\documentclass[prb,aps,showpacs,preprintnumbers,amsmath,amssymb,twocolumn,superscriptaddress,10pt]{revtex4-2}

\usepackage{graphicx}  
\usepackage{subfigure}
\usepackage{dcolumn}
\usepackage{bm}
\usepackage{amssymb}
\usepackage{tabularx}
\usepackage{threeparttable}
\usepackage{caption,booktabs}
\usepackage{xcolor}
\usepackage{multirow}
\usepackage[T1]{fontenc}
\usepackage[colorlinks,linkcolor=blue,citecolor=blue,anchorcolor=blue]{hyperref}
\begin{document}
\title{Exchange interactions and finite-temperature magnetism in (111)-oriented (LaMnO$_3$)$_{2n}$|(SrMnO$_3$)$_n$ superlattices
}
\author{Shivalika Sharma}
\email{shivalika@kentech.ac.kr}
\affiliation{Institute of Physics, Nicolaus Copernicus University, 87-100 Toru\'n, Poland}
\author{Julio do Nascimento}
\affiliation{School of Physics, Engineering and Technology, University of York, York YO10 5DD, UK}
\author{Imran Ahamed}
\affiliation{Institute of Physics, Nicolaus Copernicus University, 87-100 Toru\'n, Poland}
\author{Fabrizio Cossu}
\affiliation{School of Physics, Engineering and Technology, University of York, York YO10 5DD, UK}
\affiliation{Department of Physics and Institute of Quantum Convergence and Technology, Kangwon National University, Chuncheon, 24341, Republic of Korea}
\author{Heung-Sik Kim}
\affiliation{Department of Energy Technology, Korea Institute of Energy Technology (KENTECH), Naju, South Korea}
\affiliation{Department of Physics and Institute of Quantum Convergence and Technology, Kangwon National University, Chuncheon, 24341, Republic of Korea}
\author{Igor {Di Marco}}
\email{igor.dimarco@physics.uu.se}
\email{igor.dimarco@umk.pl}
\affiliation{Institute of Physics, Nicolaus Copernicus University, 87-100 Toru\'n, Poland}\affiliation{Department of Physics and Astronomy, Uppsala University, Uppsala 751 20, Sweden}
\date{\today}
\begin{abstract}
We present a first-principles investigation of magnetic exchange interactions and critical behavior in (111)-oriented (LaMnO$_3$)$_{2n}$|(SrMnO$_3$)$_n$ superlattices for $n=2,4,6$.
For all superlattices under investigation, we find robust half-metallic ferromagnetism extending across all the layers of both component regions. Changing octahedral tilt patterns is found to have negligible effects on the magnetic properties, despite determining the presence or absence of small Jahn-Teller distortions. The analysis of the response of the magnetic coupling to a variation of the Coulomb interaction parameters demonstrates that ferromagnetism is driven by a double-exchange mechanism involving itinerant $e_g$ electrons, while its final strength is hampered by antiferromagnetic contributions due to the superexchange of localized $t_{2g}$ electrons. Multi-scale simulations based on atomistic spin dynamics show that the thinnest superlattices, $n=2,4$, possess an ordering temperature that is at least comparable to that of La$_{2/3}$Sr$_{1/3}$MnO$_3$. Conversely, as thickness increases, a two-phase behavior emerges, where the SrMnO$_3$ region loses long-range order faster than the LaMnO$_3$ region.
While the global ordering temperature increases together with thickness, we argue that the high-temperature regime for the observed two-phase behavior is not representative of the real physical system, which will undergo a combined electronic, magnetic and structural phase transition as soon as the long-range order is lost inside the SrMnO$_3$ region.
This study provides insights into the emergent magnetic phases and transition temperatures relevant to oxide heterostructures.
\end{abstract}
\maketitle
\section{Introduction}
\noindent
Perovskite oxide superlattices have emerged as a rich platform for the realization of novel correlated electronic and magnetic phases by exploiting the interaction between dimensional confinement, interface effects, and lattice distortions~\cite{Adamo2009,Burgy2001,Dagotto2005,Imada1998,Nakao2015}. In particular,  bulk LaMnO$_3$ (LMO) and SrMnO$_3$ (SMO) are prototypical manganites with contrasting bulk properties: LMO is an A-type antiferromagnetic (AFM) Mott insulator stabilized by cooperative Jahn-Teller distortions, while SMO is a G-type AFM band insulator with negligible octahedral tilts~\cite{schmitt-PRB.101.214304,Zhu2020SrMnO3}. Their combination in heterostructures provides a charge-balanced Mn$^{3+}$/Mn$^{4+}$ framework, enabling emergent phases that are not present in either of the parent compounds~\cite{hellman2017rmp}. For magnetism specifically, the presence of Mn atoms in different ionization states favors a ferromagnetic (FM) coupling due to the double-exchange mechanism, which then competes with AFM coupling due to the superexchange mechanism that dominates in the bulk~\cite{bhattacharya-AnnuRevMR2014}. The competition between FM and AFM coupling can be tuned through layer thickness, epitaxial strain, and orientation of growth, making LMO/SMO heterostructures a flexbile playground for exploring a variety of correlated electron phenomena~\cite{KHOMSKII202498}. Most theoretical and experimental investigations to date have focused on (001)-oriented (LMO)$_{m}$|(SMO)$_n$ superlattices, where the component ratio is often adjusted to $m=2n$ to have the same stoichiometry as in the well-known ferromagnetic alloy La$_{2/3}$Sr$_{1/3}$MnO$_3$ (LSMO)~\cite{smadici2007,bhattacharya2008,adamo2008,may2008,nanda2009,smadici2012,zhang2012,Nakao2015,keunecke2020}. 
For thin superlattices, robust ferromagnetism is observed, owning to a cation distribution that is rather similar to LSMO. However, electronic structure calculations reveal that the FM coupling remains limited to the interface (IF) and therefore AFM interactions are quickly recovered when $n$ is increased~\cite{Nanda2008,nanda2009}.
Recent experimental advances have made it possible to measure the strength of the ferromagnetism at the IF upon varying thickness as well as to correlate such a strength to the extension of the measured charge transfer~\cite{keunecke2020,Curie_001,schuler2025}.
These findings highlight the tunability and technological promise of (001)-oriented manganite superlattices, but the confinement of the FM coupling to the IF may limit their practical utility in spintronic devices~\cite{dietl_book,pulizzi2012natmat11_367,chumak15natphys11_453,tanaka20jjap60_010101}.

Recently, oxide heterostructures grown along the (111) crystallographic direction have attracted considerable interest as the reduced symmetry and altered octahedral geometry can potentially host exotic correlated phases, such as quantum anomalous Hall states and other topological magnetic phases~\cite{Chandra2017,Chakhalian2020}.
While (111)-oriented oxide heterostructures are more challenging to synthesize due to substrate compatibility and growth constraint, recent advances in materials fabrication and characterization have made them accessible experimentally~\cite{Chakhalian2020,zhou2020,xu2023,wang2022pss,jansen2024prm}. 
Theoretical modeling and first-principles calculations are invaluable to investigate the physical properties of this class of systems before experimental analysis, with the double goal of identifying the most interesting regimes of stacking and composition as well as the most favorable conditions for growth. In two previous works, we employed \textit{ab-initio} electronic structure calculations to demonstrate that (111)-oriented (LMO)$_{2n}$|(SMO)$_n$ superlattices with $n=2,4,6$ possess an intriguing FM ground-state that is half-metallic across all the composing layers~\cite{Cossu2022,Cossu2024}. This behavior indicates that the FM coupling is not merely localized at the IF, but originates from a bulk-like phase transition of LMO, due to the concomitant action of stacking, epitaxial strain and charge transfer~\cite{Cossu2022}. Breathing modes and Jahn-Teller distortions may also appear, depending on magnetic order, perioditicy and tilts of the oxygen octahedra around the Mn ions~\cite{Cossu2024}. 
These results highlight crystallographic orientation as a powerful, independent design parameter for tuning exchange interactions and magnetic order, but they also leave several key questions unanswered. What is the microscopic exchange mechanism that drives the ferromagnetism? How does the Curie temperature depend on superlattice thickness? Addressing these issues is important not only for a fundamental understanding of magnetism and strong correlations in the (111) stacking geometry, but also for identifying optimal growth conditions and assessing the prospects of these superlattices for technological applications.

In the present study, we investigate the finite-temperature magnetic properties of (111)-oriented (LMO)$_{2n}$|(SMO)$_n$ superlattices with $n=2,4,6$. We employ \textit{ab-initio} electronic structure calculations to evaluate the interatomic exchange parameters and understand their dependence on superlattice thickness and tilting system. These data are then used to perform multi-scale modeling of the magnetic properties via atomistic spin dynamics (ASD) simulations, which allows us to determine ordering temperatures and to investigate the phase behavior of the different component regions. The paper is organized as follows. Section~\ref{sec:methodology} describes the methodology employed in our work, while Section~\ref{sec:magnetic exchange interaction} illustrates the trends of the magnetic coupling with respect to thickness and tilt systems. The analysis of the nature of the exchange coupling is covered in Section~\ref{sec:nature}, while the analysis of the finite-temperature magnetic properties is reported in Section~\ref{sec:crtical_temp}. Finally, the conclusions of our study and outlook for future research are reported in Section~\ref{sec:conclusions}.

\section{Methods and models}
\label{sec:methodology}
 Electronic structure calculations were performed in density-functional theory (DFT) by means of the full-potential linear muffin-tin orbital method (FP-LMTO) as implemented in the RSPt code~\cite{RSPt}. The exchange-correlation functional was treated in the generalized gradient approximation as parameterized by  Perdew-Burke-Ernzerhof (PBE-GGA)~\cite{paw,perdew_PhysRevLett.78.1396_1997,pw92}. 
   The muffin-tin spheres, which are used to partition the physical space in full-potential methods, were determined to optimize the quality of the basis set as 1.8, 2.6, 2.5 and 1.75 {a.u.} for Mn, Sr, La and O, respectively. The valence electrons were distributed over two energy sets (valence and semi-core) and accounted for $3s$, $3p$, $4s$, $4p$ and $3d$ states for Mn, $4s$, $4p$, $5s$ and $5p$ states for Sr, $5s$, $5p$, $6s$, $6p$ and $5d$ states for La, $2s$ and $2p$ states for O. 
   Two energy tails were used to describe the basis functions in the interstitial region, corresponding to kinetic energies of -0.3 Ry and -1.5 Ry. The $\mathbf{k}$-integration was performed via thermal smearing, with an effective temperature of 189~K.
   The strong Coulomb repulsion affecting the Mn-$3d$ states in manganites~\cite{KHOMSKII202498} was treated at the Hartree-Fock level by means of the DFT+U approach~\cite{Vladimir_1997}. In RSPt, this approach employs a fully rotationally invariant implementation based on a four-index U-matrix~\cite{grechnev07prb,granas12cms}, which is in turn constructed from the Slater integrals with fixed atomic ratios~\cite{Vladimir_1997,grechnev07prb}. 
   The Hubbard and Hund interaction parameters were initially set to $U=3.8$ eV and $J=1.0$ eV respectively, in agreement with previous literature~\cite{Nanda2008,Nanda2010,Cossu2022,Cossu2024}. These values were not only found to reproduce the bulk properties to a good accuracy, but are also in very good agreement with parameter-free calculations based on the meta-GGA exchange-correlation functional SCAN~\cite{Cossu2022}.
   Variations of the Coulomb interaction parameters were also investigated to individuate the contribution of the competing exchange mechanisms. Finally, the fully-localized limit (FLL) scheme was applied for the double-counting correction~\cite{Vladimir_1997}. 

 Following the self-consistent DFT+U calculations, the interatomic exchange parameters $J_{ij}$ were determined by mapping the magnetic excitations onto an effective Heisenberg Hamiltonian~\cite{szilva23rmp95_035004}:
\begin{equation}
 {H} = - \sum_{i\neq j} J_{ij}\vec{e_{i}}\cdot\vec{e_{j}}
 \label{eqn1}
\end{equation}
Here, $(i,j)$ are site indices, while $\vec{e_i}$ and $\vec{e_j}$ are unit vectors along the spin direction at sites $i$ and $j$, respectively. The $J_{ij}$ are calculated using a generalized magnetic force theorem~\cite{Igor_PhysRevB_2015,KvashninPhysRevB_2020}. The localized basis used in DFT+U as well as for the calculation of the $J_{ij}$ was constructed from the so-called muffin-tin heads as described in previous works~\cite{grechnev07prb,granas12cms,Igor_PhysRevB_2015}.

The calculated exchange interactions, $J_{ij}$, were used as input parameters for ASD simulations based on the classical Heisenberg model~\cite{book_olle}. These simulations were performed using the VAMPIRE software package~\cite{Evans_2014} with the primary goal of determining the magnetic ordering temperature (Curie temperature, $T_C$) of the systems. The calculations utilized the Metropolis Monte-Carlo algorithm to simulate the temperature dependence of the magnetization ($M$-$T$ curves)~\cite{book_olle}. For systems exhibiting a single magnetic transition, such as the superlattices with $n=2$ and $n=4$, the Curie temperature was obtained by fitting the magnetization curve to the standard  power law~\cite{landau1980statistical}:
\begin{equation}
  M(T) \propto \left(1-\frac{T}{T_{c}}\right)^\beta
  \label{eq:CurieLaw}
\end{equation}
where $\beta$ is the critical exponent.
For thicker superlattices, where the presence of distinct magnetic sublattices led to complex, multi-stage ordering, a simple power-law fit was found to be insufficient to capture the interaction between IF and the rest. In these cases, the critical temperatures were identified by calculating the macroscopic magnetic susceptibility, $\chi$. The peaks in the susceptibility $\chi(T)$ correspond to the critical fluctuations associated with the phase transitions, allowing for the determination of the critical temperatures for the different magnetic subsystems without imposing an pre-established fitting function~\cite{Binder1992}.
The simulation domains were defined as $N_x \times N_y \times N_z$ grids. For the $n=2$ and $n=4$ systems, a cubic lattice of $14 \times 14 \times 14$ unit cells was employed. To adequately capture the spatial inhomogeneities expected in the thicker $n=6$ heterostructure, a larger, anisotropic volume of $30 \times 50 \times 5$ was adopted. At each temperature step, the system was initialized from the final state of the previous increment and allowed to thermalize for 5,000 Monte Carlo steps (MCS). Following equilibration, thermodynamic observables were averaged over a data collection window of 10,000 MCS for the $n=2$ and $n=4$ systems. In contrast, the $n=6$ simulations were run with an extended sampling period of 240,000 MCS to accurately resolve the complex multi-peak structure observed in the susceptibility.

\section{Zero-temperature magnetic properties}
\label{sec:magnetic exchange interaction}
\begin{figure}
    \centering
    \includegraphics[width=0.87\linewidth]{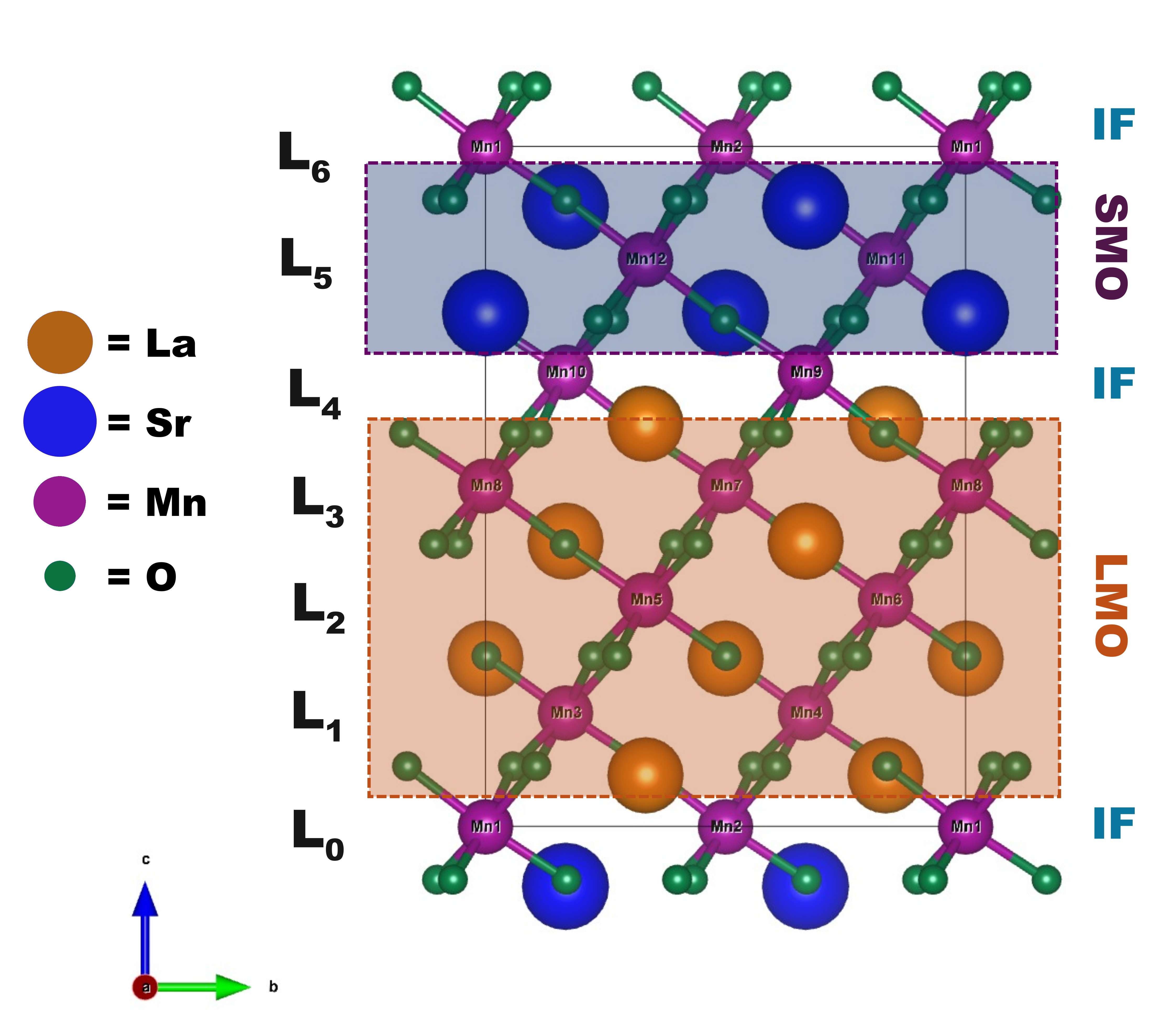}
    \caption{Side view of the (111)-oriented (LMO)$_{4}$|(SMO)$_2$ superlattice, emphasizing the stacking sequence of the various layers.}
    \label{fig:structure}
\end{figure}
The starting point of our study is the determination of the magnetic ground state of (111)-oriented (LMO)$_{2n}$|(SMO)$_n$ superlattices with $n=2,4,6$. 
The peculiarity of the (111) orientation can be inspected from the side view of the superlattice, shown in Figure~\ref{fig:structure} for $n=2$. The two most favorable arrangements of the octahedral tilts for this geometry are $a^-a^-a^-$ and $a^-a^-c^+$ (in Glazer's notation~\cite{glazer-AC:B1972,glazer-AC:A1975}), which correspond to the space groups $R\bar{3}c$ and $Pnma$. 
In our previous study~\cite{Cossu2024}, we demonstrated that full structural relaxation in DFT+U with PBE-GGA leads to a half-metallic FM ground state for $n=2,4,6$ in both tilting systems. However, for $n=2$, no stable solution exists for $a^-a^-c^+$, independently from the magnetic order, without involving a substantial epitaxial strain. The analysis of the magnetic coupling will thus focus on the three FM ground states obtained for the $a^-a^-a^-$ tilting system with $n=2,4,6$ and on the FM ground state obtained for the $a^-a^-c^+$ tilting system with $n=4$. These states are sufficient to provide an overview of the role played by thickness and tilting system. 

The optimized structures reported in our previous study~\cite{Cossu2024} and obtained by means of the Vienna \textit{ab initio} Simulation Package (VASP)~\cite{vasp1,vasp2,vasp3} were used to perform DFT+U calculations via RSPt, as described in Section~\ref{sec:methodology}.
The Brillouin zone was sampled with a dense Monkhorst-Pack grid of $11\times7\times2$, $11\times7\times1$ and $11\times7\times1$ {\bf{k}}-points for $n=2,4,6$ respectively.
\begin{figure*}
    \centering
    \includegraphics[width=1\linewidth]{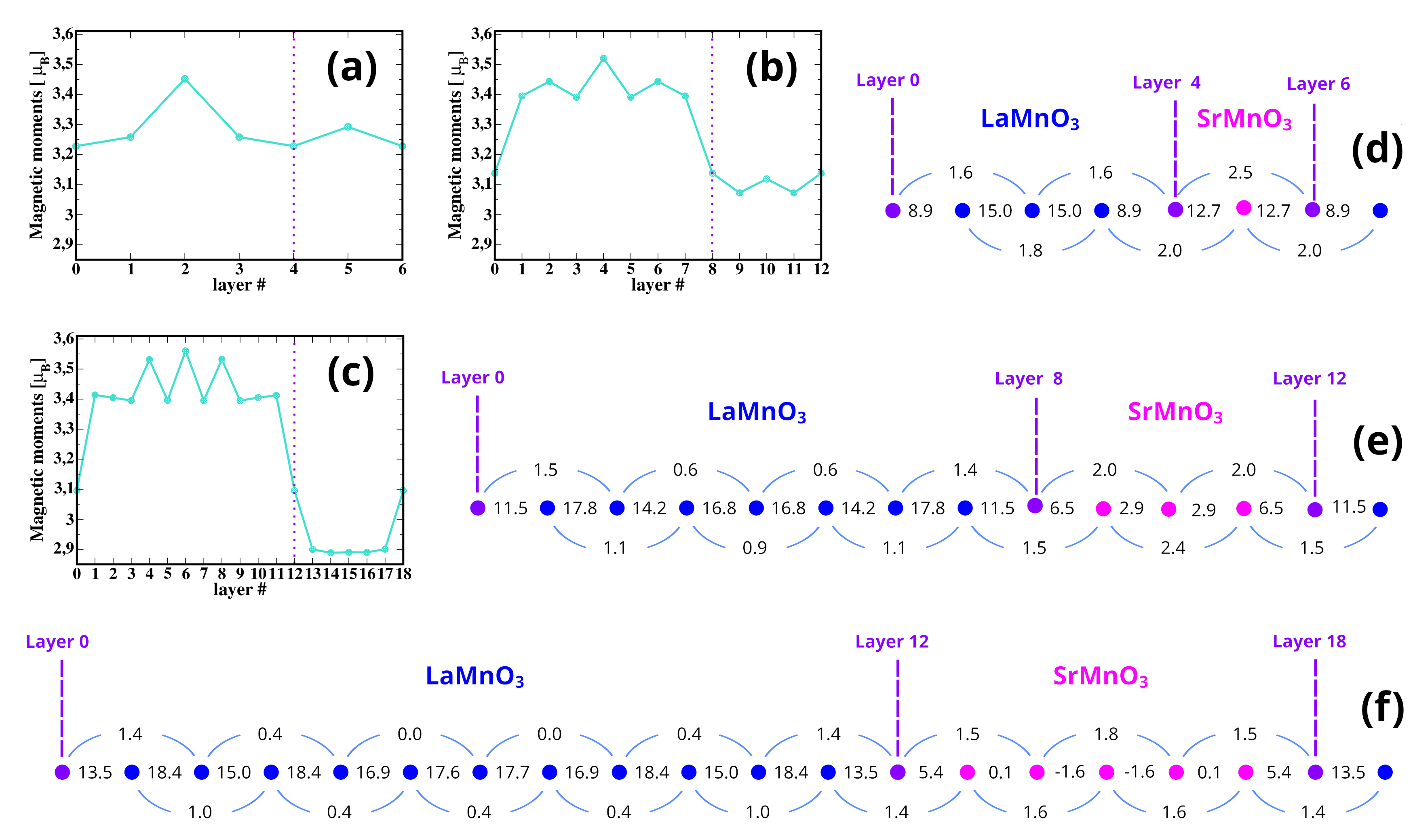}
    \caption{Layer-resolved magnetic moments at the Mn sites in the (111)-oriented (LMO)$_{2n}$|(SMO)$_n$ superlattices in the $a^-a^-a^-$ tilting system for $n=2,4,6$ (a-c). Corresponding layer-resolved interatomic exchange interactions $J_{ij}$ (in meV) between the magnetic moments at the Mn sites for the first (straight segments) and fourth (semi-circular arcs) nearest neighbors along the (001) direction (d-f). The data for $n=6$ have already been reported in Ref.~\onlinecite{Cossu2022} and are shown here for an easier comparison.}
    \label{fig:momjija-a-a-}
\end{figure*}
The layer-resolved profile of the spin magnetic moments at the Mn sites of (LMO)$_{2n}$|(SMO)$_n$ superlattices in the $a^-a^-a^-$ tilting system is reported in Figure~\ref{fig:momjija-a-a-}(a-c). The three investigated thicknesses show various common features, in agreement with our previous study~\cite{Cossu2024}. First, the magnetic moments oscillate across all the layers of the superlattice, although a flatter pattern is observed inside the SMO region for $n=6$. These oscillations are coupled to the breathing distortions of the oxygen octahedra~\cite{Cossu2022,Cossu2024}. Notice that in this tilting system, the typical Jahn-Teller distortions arising in LMO are quenched by symmetry~\cite{varignon19prb,varignon19prr}. Second, the maximum (minimum) value of the magnetic moments is reached in Mn layers deep inside the LMO (SMO) region. The layers at the IF have an intermediate behavior, in stark contrast with the results obtained for the (001) orientation~\cite{Nanda2008,nanda2009,Nanda2010}. This suggests that the mechanism driving the FM order is due to a phase transition happening inside the LMO region and is not exclusively due to the charge transfer across the IF~\cite{Cossu2022}. This is confirmed by the spread between maximum and minimum values inside each superlattice, which seems to increase with thickness. This trend is the first hint of a competition between the LMO and SMO regions in driving the magnetic order of the superlattice.

A more detailed insight can be obtained through the inspection of the interatomic exchange parameters. The calculated $J_{ij}$ show a strong dependence on exchange-path connectivity and bond geometry, as typical of magnetic oxides~\cite{dionnebook}. For a Mn atom at site $i$, the largest terms arise for sites $j$ corresponding to its first nearest neighbors, which are three Mn atoms in the layer above and three Mn atoms in the layer below with respect to the direction of growth (see Figure~\ref{fig:structure}). The couplings with the second and third nearest neighbors are totally suppressed due to unfavorable bond geometry, in agreement with the Goodenough-Kanamori-Anderson (GKA) rules~\cite{Goodenough55,Kanamori1959,Anderson1959, Goodenough1963}. The coupling with the fourth nearest neighbors, which are found along a straight Mn--O--Mn--O--Mn line (see again Figure~\ref{fig:structure}) is finite, but small. Farther atoms show negligible interactions with respect to the first and fourth neighbors. The symmetry of the $R\bar{3}c$ structure dictates that the couplings depend only on the layer. Therefore, it is convenient to visualize the trends of the exchange interaction between the first and fourth Mn nearest neighbors via the chain-like schemes shown in Figure~\ref{fig:momjija-a-a-}(d-f). 
The exchange interactions are mostly FM (positive sign) with a few AFM (negative sign) terms for $n=6$. 
For all compositions, the first nearest neighbor exchange is largest inside the LMO region and smallest inside the SMO region. This is in agreement with the data on the magnetic moments and is a direct consequence of the definition of the effective Heisenberg Hamiltonian in Eq.~\ref{eqn1}. 
The maximum values tend to increase gradually with $n$, going from 15 meV for $n=2$ to 17.4 meV and 18.4 meV for $n=4$ and $n=6$, respectively. 
\begin{figure*}[t]
    \centering
    \includegraphics[width=1\linewidth]{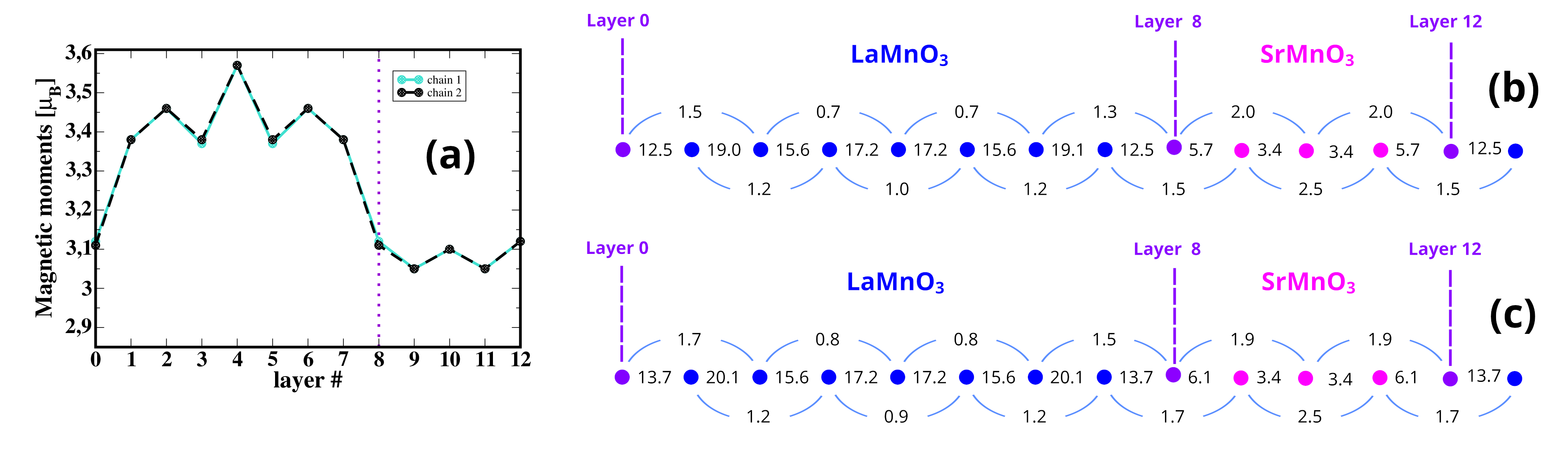}
    \caption{Layer-resolved magnetic moments at the Mn sites in the (111)-oriented (LMO)$_{8}$|(SMO)$_4$ superlattice in the $a^-a^-c^+$ tilting system, separated in two distinct chains arising from the asymmetric tilt pattern (a). Corresponding layer-resolved interatomic exchange interactions $J_{ij}$ (in meV) between the magnetic moments at the Mn sites for the first (straight segments) and fourth (semi-circular arcs) nearest neighbors along the (001) direction (b-c).}
    \label{fig:momjija-a-c+}
\end{figure*}
No AFM interaction is observed within the LMO layers, no matter the thickness. On the contrary, the strength of the FM coupling seems to increase with the extension of the LMO region. This is markedly different from the results obtained for (001)-oriented superlattices, where an AFM coupling accompanied by a locally insulating character is recovered when the LMO region becomes thicker than a couple of layers~\cite{Nanda2008,nanda2009,Nanda2010}. This behavior is counterintuitive, as one would expect to recover the bulk-like properties of LMO at large $n$. The observed discrepancy between our calculations and our expectations will be explained when we will illustrate the finite-temperature magnetic properties, in Section~\ref{sec:crtical_temp}. 
In the SMO region, we observe a gradual crossover from FM to AFM coupling for the first nearest neighbors as the layer thickness increases from $n=2$ to $n=6$. This trend can be explained with the strain-induced suppression of the intrinsic AFM G-type order of bulk SrMnO$_3$, where epitaxial strain can tune the balance for an AFM-FM transition~\cite{Ricca2019SrMnO3,Zhu2020SrMnO3}. This thickness-dependent weakening of the FM coupling is also observed at the IF layers. 
When looking at the coupling between fourth nearest neighbors in Figure~\ref{fig:momjija-a-a-}(d-f), we note that the values are significantly smaller and seem to decrease with thickness. An interesting feature is that for the largest superlattice, $n=6$, these terms are responsible for the stabilization of the FM order inside the SMO region. As we will see in Section~\ref{sec:crtical_temp}, these competing interactions lead to interesting properties on the finite-temperature magnetism.

Overall, the data reported in Figure~\ref{fig:momjija-a-a-} demonstrate that the superlattice for $n=2$ has a behavior that resembles that of LSMO, with an almost uniform distribution of magnetic moments and exchange interactions. The strength of the first nearest-neighbor $J_{ij}$ is 30\% to 115\% higher than the corresponding term in LSMO, depending on the layer~\cite{Bottcher_2013}. Conversely, the strength of the fourth nearest-neighbor $J_{ij}$ is 20\% to 35\% smaller than the corresponding term in LSMO,  depending on the layer~\cite{Bottcher_2013}.
An overall stronger effective coupling in the ordered superlattice than in a compositionally equivalent random alloy is plausible, since the superlattice suppresses chemical disorder and associated bond-geometry fluctuations, thereby enhancing the coherence of the relevant exchange pathways.
However, we cannot discard the possibility that our methods are overstimating the strenght of the magnetic coupling, as we will also discuss in the Conclusions. 
In relative terms, we notice that increasing thickness makes the FM coupling inside the LMO region even stronger~\footnote{In our previous study~\cite{Cossu2022}, we made a different claim because we missed a prefactor when converting the $J_{ij}$ from the Heisenberg Hamiltonian used in Ref.~\cite{Bottcher_2013}.}.
This does not imply that the system with largest thickness has better magnetic properties, though, due to the weakly coupled SMO region.

Changing the tilting system to $a^-a^-c^+$ leads to the $Pnma$ structure, which has twice as many atoms as the  $R\bar{3}c$ structure. This doubling of the periodicity splits the Mn atoms into two nonequivalent sublattices that can be conveniently visualized as two Mn chains~\cite{Cossu2024}.
Although these two Mn chains form due to the inherent asymmetry of  $Pnma$ structure, the itinerant character of the electronic system mitigates the differences between sublattices. This can be verified by the inspection of the layer-resolved magnetic moments and interatomic exchange interactions, reported in Figure~\ref{fig:momjija-a-c+}, for $n=4$. The comparison with the $a^-a^-a^-$ tilting system reveals no qualitative change in the trends across the layers of the superlattice. Quantitatively, the additional symmetry-breaking seems to increase the values of the Mn magnetic moments and corresponding exchange interactions, as expected from DFT+U using FLL~\cite{peters14prb89_205109}. This increase may also be connected to the fact that the Jahn-Teller distortions are no longer quenched in the $a^-a^-c^+$ tilting system, albeit they remain smaller than the breathing distortions~\cite{Cossu2024}.

\section{Nature of the exchange coupling}
\label{sec:nature}
In Section~\ref{sec:magnetic exchange interaction}, we illustrated the trends of the magnetic properties in the investigated superlattices with respect to thickness and tilting system. However, a limited insight was provided on the origin of the FM coupling driving the long-range order.
To better understand this aspect, we focus on the $n=2$ superlattice with the $a^-a^-a^-$ octahedral tilt pattern, which is visualized in Figure~\ref{fig:structure}. We select Mn atoms in layers 2 (inside LMO), 5 (inside SMO) and 4 (at the IF) as representative of different regions.
For those atoms, we compute the orbital decomposition of the exchange interactions $J_{ij}$ on the basis of the octahedral symmetry~\cite{kvashnin16prl116,cardias17scirep}. The corresponding values of the  $E_g$-$E_g$, $E_g$-$T_{2g}$ and $T_{2g}$-$T_{2g}$ components for the first and fourth nearest neighbors are reported in Table~\ref{tab:orbres}.
For the Mn atom sitting at the IF there are two values, corresponding to the coupling with Mn atoms inside the LMO and SMO regions, and thus labeled as IF$_\text{L}$ and IF$_\text{S}$, respectively.
Before discussing the details of the orbital decomposition, we note that the traced values reported in Table~\ref{tab:orbres} are sensibly smaller than those reported in Figure~\ref{fig:momjija-a-a-}(d). This is due to the fact that the $J_{ij}$ discussed above were obtained by employing generalized L\"owdin projectors to evaluate the intersite Green's functions~\cite{Igor_PhysRevB_2015}, while the orbital decomposition requires atomic-like projectors to retain the proper symmetry~\cite{kvashnin16prl116,cardias17scirep} 
Usually, these two approaches give very similar results but in our calculations we found a difference of more than a factor 2. For technical reasons related to the completeness of the basis, we are confident that the numbers provided in Figures~\ref{fig:momjija-a-a-} and~\ref{fig:momjija-a-c+} are quantitatively more accurate, but we can still use the the values reported in Table~\ref{tab:orbres} for a qualitative analysis of the relative contribution to the total exchange.


Thus, we move to the analysis of the orbital-decomposed values in Table~\ref{tab:orbres}. For all three regions, the first nearest-neighbor terms are dominated by a strongly FM $E_g$-$E_g$ contribution, followed by smaller AFM contributions from the   $T_{2g}$-$T_{2g}$ and $E_g$-$T_{2g}$ channels. The fourth nearest-neighbor exchange is significantly  weaker, but is still dominated by FM contributions stemming from the $E_g$-$E_g$ channel. The $T_{2g}$-$T_{2g}$ terms decrease of two orders of magnitude with respect to the first nearest neighbors, which suggests an exchange mechanism involving localized electrons for this symmetry.
\begin{table}[b]
\caption{Orbital-resolved interatomic exchange interactions between magnetic moments at the Mn sites in the (111)-oriented (LMO)$_{4}$|(SMO)$_2$ superlattice in the $a^-a^-a^-$ tilting system. First and fourth nearest-neighbor interactions are shown for Mn atoms inside LMO, SMO and IF regions. These Mn atoms belong to the layers 2, 5 and 4 depicted in Figure~\ref{fig:structure}, respectively. For the Mn atoms at the IF two values are reported, namely IF$_\text{L}$ and IF$_\text{S}$, corresponding to coupling to Mn atoms belonging to the LMO and SMO regions, respectively (see also Figure~\ref{fig:structure}). Due to the symmetry of the IF for $n=2$, the first (fourth) nearest-neighbor terms of IF$_\text{S}$ and SMO (IF$_\text{L}$ and LMO) should be equal. Their discrepancy is a measure of the numerical precision achieved in our calculations.
Data obtained for three different values of the Hubbard $U$ are reported. Notice that these $J_{ij}$ are calculated with a different projection scheme with respect to those reported in Figures~\ref{fig:momjija-a-a-} and~\ref{fig:momjija-a-c+}, as explained in the main text.}
\label{tab:orbres}
\centering
\renewcommand{\arraystretch}{1.15}
\begin{tabular}{|c|c|c|r|r|r|r|}
\hline
\hline
\multirow{2}{*}{NN} &
\multirow{2}{*}{U(eV)} &
\multirow{2}{*}{Region} &
\multicolumn{3}{c|}{\textbf{Decomposition (meV)}} &
\multirow{2}{*}{\textbf{$J_{ij}$(meV)}} \\
\cline{4-6}
 & & &
 E$_g$-E$_g$ &
 T$_{2g}$-T$_{2g}$ &
 E$_g$-T$_{2g}$ &
 \\
\hline

\multirow{12}{*}{1st} &
\multirow{4}{*}{3.8} &
LMO &  \text{13.59} & \text{-4.77} & \text{-0.81} & \text{8.01} \\
& &SMO & \text{13.10} & \text{-7.11} & \text{-0.27} & \text{5.72} \\
& &IF$_\text{L}$  & \text{8.40} & \text{-4.30} & \text{-0.38} & \text{3.72} \\
& &IF$_\text{S}$  & \text{13.05} & \text{-7.11} & \text{-0.27} & \text{5.67} \\
\cline{2-7}
& \multirow{4}{*}{5.0} &
LMO & \text{17.71} & \text{-4.29} & \text{-0.74} & \text{12.68} \\
& & SMO& \text{17.38} & \text{-6.36} & \text{-0.19} & \text{10.83} \\
& & IF$_\text{L}$& \text{11.04} & \text{-3.85} & \text{-0.22} & \text{6.97} \\
& &IF$_\text{S}$  & \text{17.32} & \text{-6.36} & \text{-0.19} & \text{10.77}\\ 
\cline{2-7}
& \multirow{4}{*}{7.0} &
LMO &   \text{17.21} & \text{-2.82} & \text{1.90} & \text{16.29} \\
& &SMO & \text{18.54} & \text{-4.58} & \text{1.25} & \text{15.21} \\
& &IF$_\text{L}$  & \text{12.90} & \text{-2.57} & \text{1.88} & \text{12.21} \\
& &IF$_\text{S}$  & \text{18.59} & \text{-4.58} & \text{1.24} & \text{15.25} \\
\hline

\multirow{12}{*}{4th} &
\multirow{4}{*}{3.8} &
LMO &  \text{1.11} & \text{0.04} & \text{-0.16} & \text{0.99} \\
& &SMO & \text{1.34} & \text{0.06} & \text{-0.16} & \text{1.24} \\
&&IF$_\text{L}$  & \text{1.11} & \text{0.05} & \text{-0.16} & \text{0.99}\\
&&IF$_\text{S}$  & \text{1.57} & \text{0.08} & \text{-0.06} & \text{1.59}\\
\cline{2-7}
& \multirow{4}{*}{5.0} &
LMO & \text{1.36} & \text{0.03} & \text{-0.15} & \text{1.24} \\
& &SMO& \text{1.65} & \text{0.04} & \text{-0.14} & \text{1.55} \\
& &IF$_\text{L}$& \text{1.36} & \text{0.03} & \text{-0.15} & \text{1.24} \\
& &IF$_\text{S}$& \text{1.97} & \text{0.05} & \text{-0.06} & \text{1.96} \\
\cline{2-7}
& \multirow{4}{*}{7.0} &
LMO & \text{1.36} & \text{0.01} & \text{-0.12} & \text{1.25} \\
& & SMO& \text{1.52} & \text{0.01} & \text{0.16} & \text{1.69} \\
& & IF$_\text{L}$& \text{1.36} & \text{0.01} & \text{-0.12} & \text{1.26} \\
& & IF$_\text{S}$& \text{2.27} & \text{0.03} & \text{-0.06} & \text{2.24} \\
\hline
\end{tabular}
\end{table}

\begin{figure}[t]
\centering
\includegraphics[width=0.85\linewidth]{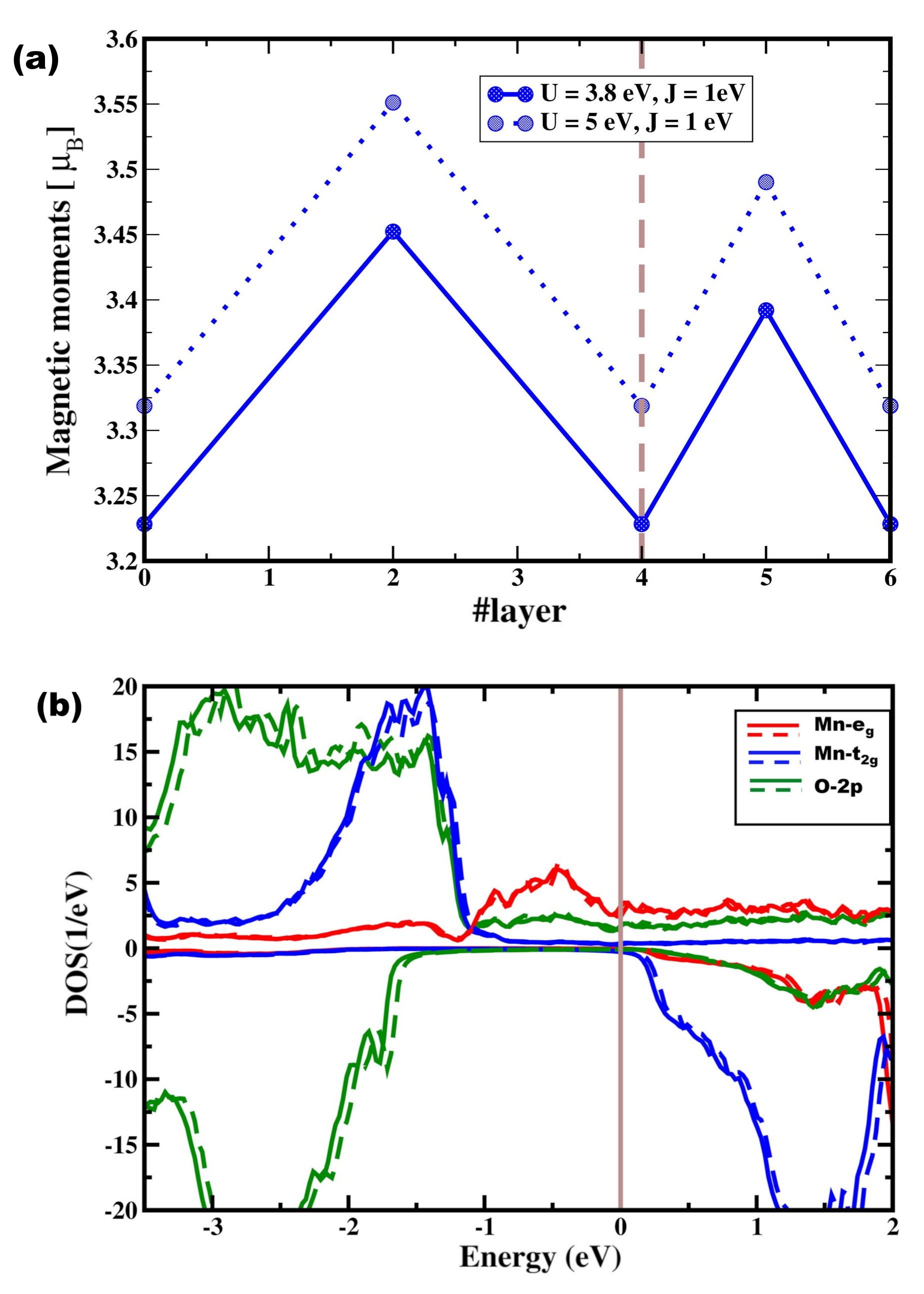}
\caption{Layer-resolved magnetic moments at the Mn sites in the (111)-oriented (LMO)$_{4}$|(SMO)$_2$ superlattice in the $a^-a^-a^-$ tilting system, for two different sets of Coulomb interaction parameters (a). Layer-integrated Mn-$3d$ and O-$2p$ PDOS as obtained for $U = 3.8$~eV (solid lines) and $U = 5.0$~eV (dashed lines), with a fixed $J = 1$~eV (b).}
\label{fig:magdosn2}
\end{figure}

The behavior of the orbital-decomposed exchange interactions can be understood from the details of the electronic structure. The projected density of states (PDOS) for the Mn-$3d$ and O-$2p$ states is shown in Figure~\ref{fig:magdosn2}(b).
Partially occupied itinerant Mn-$3d$ states with $E_g$ symmetry are present near the Fermi level, overlapping with a finite O-$2p$ contribution. The metallic character of the $e_g$ states and their hybridization with the O-$2p$ states suggest a FM double-exchange mechanism between $e_g$ electrons mediated by the O-$2p$ states~\cite{Zener1951,AndersonHasegawa1955_DoubleExchange}. This is not a surprise, since several theoretical and experimental studies have reported the emergence of oxygen-mediated double-exchange coupling in a variety of manganites~\cite{PhysRevB_Zhao,Park1998,DagottoHottaMoreo2001_PhysRep_Manganites}. The surprising aspect here is that the FM $E_g$-$E_g$ contribution is not strongest at the IF, but inside the LMO and SMO regions. This is counterintuitive if one focuses only on charge transfer, which is expected to be most localized near the IF. 
However, double exchange is controlled primarily by the kinetic-energy scale of the $e_g$ carriers, i.e., the effective hopping (transfer integral), rather than by the mere presence of charge transfer~\cite{degennes1960pr,tokura2000science}.
Indeed, the layer-resolved PDOS indicates that the $e_g$-derived bandwidth (roughly corresponding to the effective hopping) is largest in the innermost layers of the LMO region, decreases gradually toward the IF, and then remains approximately constant throughout the SMO region~\cite{Cossu2024}.
Then, the crucial question is why the $e_g$ electrons are less itinerant (smaller effective hopping) for the bonds corresponding to IF$_\text{L}$ than for those corresponding to SMO. 
We attribute this to a reduced Mn–O–Mn bond angle (see discussion below), which decreases the ligand-mediated $e_g$ overlap and hence renormalizes the effective hopping~\cite{tokura2000science,Coey01031999}.
Such a bond-angle reduction at the IF from the LMO side is consistent with enhanced interfacial orbital polarization and local structural/electrostatic distortions~\cite{tebano2008prl,marin2015nanolet}, which are expected to be particularly strong in the (111) geometry~\cite{Chandra2017,Chakhalian2020}. The combined reduction of $e_g$ bandwidth and Mn–O–Mn bond angle therefore makes the double-exchange contribution minimal at the IF, consistent with the traditional double-exchange models~\cite{AndersonHasegawa1955_DoubleExchange,degennes1960pr}.

\begin{figure*}[t]
\includegraphics[width=0.9\linewidth]{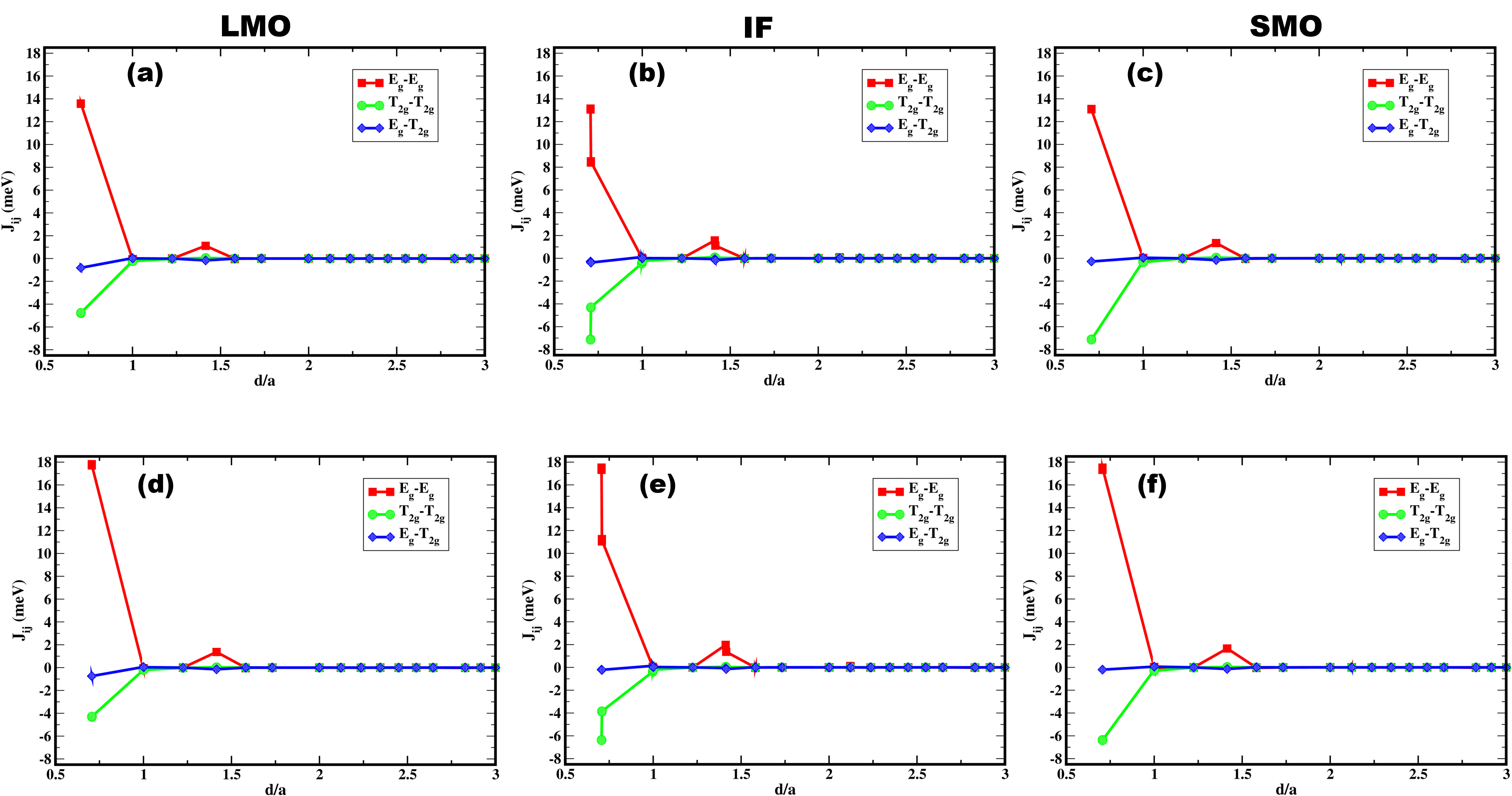}
\caption{Orbital-resolved interatomic exchange interactions $J_{ij}$ as a function of the interatomic distance $d/a$ for Mn atoms located inside the LMO and SMO regions as well as at the IF (see main text). Top panels (a-c) correspond to calculations for $U = 3.8$~eV and $J = 1$~eV, while bottom panels (d-f) correspond to calculations for $U = 5.0$~eV and $J = 1$~eV.
}
\label{fig:Jij_decomp}
\end{figure*}

From Figure~\ref{fig:magdosn2}(b), we also observe that the $t_{2g}$ states are fully localized. Thus, it is natural to conclude that the AFM $T_{2g}$-$T_{2g}$ interaction originates from the superexchange between localized half-filled $t_{2g}$ orbitals mediated by oxygen, consistently with the GKA rules~\cite{Goodenough55,Kanamori1959,Anderson1959, Goodenough1963}. This also explains the rapid quench of the $T_{2g}$-$T_{2g}$ interaction with interatomic distance that is observed in Table~\ref{tab:orbres}.
The Mn--O--Mn bond angles, shown in Table~S1 of the Supplementary Material (SM)~\cite{supplemental}, provide further insight on the strength of the super-exchange interaction across the layers. The angles of the bonds follow the general trend $\mathrm{SMO} > \mathrm{IF}_\mathrm{S} > \mathrm{IF}_\mathrm{L} > \mathrm{LMO}$,
but $\mathrm{IF}_\mathrm{L}$ and LMO are inverted for $n=2$ due to the very small thickness. Larger angles correspond to straighter Mn--O--Mn bonds and larger orbital overlap between Mn-$3d$ and O-$2p$ states, which favors more efficient virtual processes. According the GKA rules~\cite{Goodenough55,Kanamori1959,Anderson1959, Goodenough1963}, AFM super-exchange becomes strongest for angles that are closer to $180^\circ$, and thus the hierarchy above is also reflected in the $T_{2g}$-$T_{2g}$ component of the first nearest neighbor coupling in Table~\ref{tab:orbres}. Finally, we note that the larger bond angles observed inside the SMO region also contribute to the relative increase of the double-exchange $E_g$-$E_g$ term~\cite{tokura2000science,Coey01031999} with respect to the layers at the IF, as discussed in the previous paragraph.

To validate our hypothesis on the nature of the exchange coupling, we analyze the response of this system to a small variation of the Coulomb interaction parameters, by following a procedure recently applied to study  transition-metal-doped topological insulators~\cite{PhysRevB_Bi2Se3}.
Hence, we perform additional calculations where the Hubbard $U$ is increased from 3.8~eV to 5.0~eV, while the Hund's exchange is kept fixed to 1~eV. The corresponding orbital-resolved exchange parameters are reported in Table~\ref{tab:orbres}. For completeness, the dependence of the orbital-decomposed $J_{ij}$ of the selected Mn atoms on the interatomic distance for both values of $U$ is shown in Figure~\ref{fig:Jij_decomp}.

As expected, the AFM contributions due to the $T_{2g}$-$T_{2g}$ and $E_{2g}$-$T_{2g}$ channels decrease, consistently with the inverse dependence of the superexchange on the on-site Coulomb interaction, $J_{\mathrm{SE}} \sim t^2/U$~\cite{Anderson1959}. On the contrary, the FM $E_g$-$E_g$ contribution is enhanced for all layers, leading to an overall increase of the total FM exchange interaction $J_{ij}$. As shown in Figure~\ref{fig:magdosn2}(b), the Mn-$3d$ PDOS and the O-$2p$ PDOS remain almost unchanged around the Fermi energy when $U$ is increased. This indicates that the itinerant carriers mediating the ferromagnetism possess a mixed $d$-$p$ character rather than being purely Mn-$d$ derived~\cite{PhysRevB_Bi2Se3}.
Thus, the stronger Coulomb interaction does not modify the effective hopping substantially, while it still acts on the exchange splitting via the DFT+U corrections~\cite{Vladimir_1997}. In terms of the double-exchange model, this mechanism increases the energy gained from the Hund's coupling between $e_g$ electrons and localized moments formed by $t_{2g}$ electrons. This happens first because of larger magnetic moments, as visualized in Figure~\ref{fig:magdosn2}(a), and second because of a stronger effective Hund's (Kondo-like) coupling. The latter increases due to the fact that a larger $U$ increases the orbital separation between $e_g$ and $t_{2g}$ electrons, thus reducing their hybridization and suppressing screening between them. This affects the downfolding renormalizations that lead to the effective Hund's coupling, which in turn determines how much energy is gained by ferromagnetically aligned moments with respect to non-aligned ones~\cite{Zener1951,AndersonHasegawa1955_DoubleExchange}. 
This increase in the double-exchange energy corresponds to the trend that we observe for the $E_g$-$E_g$ components in Table~\ref{tab:orbres}.

\begin{figure*}[t]
    \centering
    \includegraphics[width=1\linewidth]{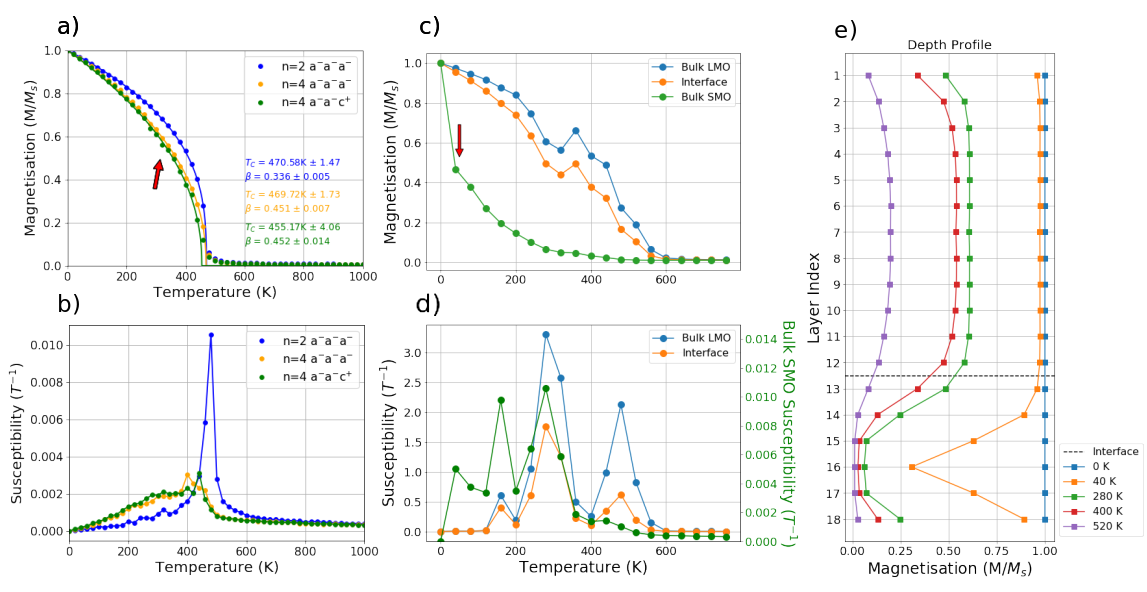}
    \caption{Finite-temperature magnetic properties of the (111)-oriented (LMO)$_{2n}$|(SMO)$_n$ superlattices. $M$-$T$ curves for $n=2,4$ (a) and corresponding magnetic susceptibility $\chi$ (b). $M$-$T$ curves for $n=6$ (c) and corresponding magnetic susceptibility $\chi$ (d), separated in relevant component regions. Magnetization depth profiles across the superlattice for $n=6$, at selected temperatures (e). Red arrows indicate the change of curvature in the $M$-$T$ curves due to the demagnetization of SMO.}
    \label{fig:CurieTemperatures}
\end{figure*}

Additional calculations were also performed for $U = 7.0$~eV, as shown in Table~\ref{tab:orbres}. A further enhancement of the FM $E_g$-$E_g$ contribution is observed, alongside a significant suppression of the AFM $T_{2g}$-$T_{2g}$ and $E_g$-$T_{2g}$ terms. This trend is consistent with the previous analysis, as one can also verify from the inspection of the PDOS of all Mn atoms (data not shown). 
The physical picture provided by these data for the (111)-oriented (LMO)$_{4}$|(SMO)$_2$ superlattice can then be used to speculate on the origin of the FM order for a larger thickness as well. In this family of superlattices, the double-exchange mechanism associated to the $E_g$-$E_g$ contribution drives the FM order, boosted by the carrier delocalization favored by the (111) stacking sequence. The AFM superexchange associated with the $T_{2g}$-$T_{2g}$ and $E_g$-$T_{2g}$ contributions remains weak for thin superlattices, overshadowed by the stronger double exchange. However, for thicker superlattices, these components become dominant inside the SMO region, favored by angles that are closer and closer to $180^\circ$, see Table~S1 in SM~\cite{supplemental}. In our calculations this happens already for $n=6$, where the first nearest-neighbor interactions inside SMO become weakly AFM. In this case, the FM order at zero temperature is stabilized by the FM double exchange term due to the fourth nearest neighbors. This order is very weak and is not expected to survive when temperature is introduced, as discussed in the next section.

\section{Critical Temperature and Magnetic Ordering}
\label{sec:crtical_temp}
We employed ASD simulations to calculate the finite-temperature magnetic properties, using the previous interatomic exchange interactions as input.
$M$-$T$ curves for superlattices with $n=2,4$ are presented in Figure~\ref{fig:CurieTemperatures}(a).
The thinnest superlattice, $n=2$, exhibits a Curie-Weiss behavior with a single magnetic transition at $T_c = 471$ K and a critical exponent $\beta \approx 0.34$. This value is in excellent agreement with the three-dimensional Heisenberg universality class ($\beta \approx 0.36$)~\cite{Campostrini2002}, indicating that short-range nearest-neighbor magnetic interactions dominate in the formation of long-range order.
In contrast, the superlattices with n = 4 show a deviation from this universality class. 
The critical exponent $\beta$ is found to be equal to about 0.45 for the $a^-a^-a^-$ tilt pattern and about 0.48 for the $a^-a^-c^+$ tilt pattern.
The deviation from the three-dimensional Heisenberg universality class originates from the competition between the two distinct regions of the heterostructure. This can be promptly verified by the inspection of the magnetic susceptibility $\chi$, shown in Figure~\ref{fig:CurieTemperatures}(b). The sharp peak observed for  $n=2$ is replaced by much broader maxima for $n=4$, almost independently from the tilt system. This feature suggests that the magnetic transition is no longer uniform; instead, different parts of the system begin to order at different temperatures, smearing out the critical singularity. Thus, the calculated $T_c$ of 455~K and 470~K, depending on the tilting system, should be interpreted with care. 

For the thickest superlattice, $n=6$, a complex multi-phase behavior that cannot be described by a standard single-Curie law emerges from the calculations. It is therefore better to separate the $M$-$T$ curves in LMO, SMO and IF regions, as illustrated in Figure~\ref{fig:CurieTemperatures}(c). We note that the hard (LMO) and soft (SMO) components do not disorder simultaneously with increasing temperature, but they rather interact as a strongly coupled thermodynamic system characterized by two distinct critical events. In our model, upon increasing temperature, the magnetization in the SMO region quickly drops but remains sustained by the strong magnetism of LMO. A first transition occurs at approximately 280~K and corresponds to SMO becoming magnetically disordered.
An anomalous susceptibility spike at the same temperature is observed in the harder LMO region, as evident from Figure~\ref{fig:CurieTemperatures}(d). Although the LMO region is well below its own ordering temperature, the sudden disordering of the adjacent SMO layers removes the exchange boundary condition at the IF. This induces spin fluctuations that propagate into the hard region, creating a ``cross-talk'' effect where the phase transition of the soft layer is imprinted onto the susceptibility of the hard layer.
Following this first transition, the system enters a magnetic proximity region in the intermediate temperature range between 280~K and 520~K. In this regime, while the deep bulk of the SMO region is disordered, the layers around the IF retain a significant non-zero induced moment, driven by the exchange field from the ordered LMO region. This behavior is better visualized in the depth profile shown in Figure~\ref{fig:CurieTemperatures}(e), where the IF acts as an exchange spring, mediating order between the two subsystems. Then, a second transition occurs at about 520~K, where the magnetic order in the LMO region also collapses. At this point, the entire superlattice becomes paramagnetic (PM). 

The results presented above highlight a fundamental crossover in the magnetic behavior as the superlattice period increases. For $n=2$, the strength of the exchange coupling, mainly defined by first and fourth nearest-neighbor shells, forces the LMO and SMO components to order collectively as a single effective medium with a unified Curie temperature of about 270~K. For $n=4$, this picture starts breaking down, as the SMO region is characterized by a weaker exchange coupling, as shown in Figures~\ref{fig:momjija-a-a-}(e) and~\ref{fig:momjija-a-c+}(b,c). 
Upon increasing temperature, the thin SMO region (4 layers) remains ordered uniquely via the coupling with the LMO/IF layers, as evidenced by the change of slope indicated by the red arrow in Figure~\ref{fig:CurieTemperatures}(a) and by the broad peaks in Figure~\ref{fig:CurieTemperatures}(b). 
The global Curie temperature predicted by our simulations for $n=4$ goes from 455~K to 470~K, depending on the tilting system. For $n=6$, a phase separation emerges unequivocally, due to the extremely weak magnetic coupling of the innermost layers in the SMO region, see Figure~\ref{fig:momjija-a-a-}(f). The system acts as composed of two semi-independent magnetic reservoirs that communicate through interfacial exchange. This decoupling allows the SMO and LMO regions to undergo separate phase transitions at 280~K and 520~K, respectively, while the IF maintains a localized proximity state that bridges these two thermodynamic regimes. 

The trend observed in the global $T_c$ points to an increase of magnetism in thicker superlattices, which is in contradiction with an intuitive understanding of the calculated magnetic coupling. These data highlight a problem of our model based on ASD simulations with parameters extracted from the magnetic ground states. In a realistic system, when temperature will make the SMO region sufficiently disordered, an electronic and structural phase transition will happen, since these degrees of freedom are strongly coupled in magnetic oxides~\cite{KHOMSKII202498,dionnebook}. As a result, the innermost layers of SMO are likely to become PM insulators, which in turn will weaken the magnetism in the rest of superlattice. Even if magnetic domains survive in the LMO regions, those domains are going to be separated by insulating PM regions and unable to ``communicate'' with each other. This will weaken the external field experienced by the SMO regions, further contributing to weaken the residual order in those layers closer to the IF. Estimating these mechanisms is not possible in our computational scheme, but something can be inferred from the $M$-$T$ curves. From Figure~\ref{fig:CurieTemperatures}(c), as indicated by the red arrow, a change of curvature is observed when the SMO order begins to collapse for $n=6$. Even by assuming that this process is delayed by the coupling with the LMO region, we doubt that the magnetic order could survive above 60~K without an an electronic-structural phase transition happening. A similar estimate can be made for $n=4$ from the noticeable change of curvature seen in Figure~\ref{fig:CurieTemperatures}(a), yielding a value of about 380~K. We can therefore conclude that even in the most negative scenarios, our theoretical model suggests that the physical properties of superlattices with $n=2,4$ make them worth of a more systematic experimental investigation, while the fragility of the magnetic order in thicker superlattices is going to be a severe obstacle for any application outside of the laboratory.

\section{Conclusions}
\label{sec:conclusions}
We have reported on our comprehensive investigation of the finite-temperature magnetic properties of (111)-oriented (LMO)$_{2n}$|(SMO)$_n$ superlattices for $n=2,4,6$. The interest in these systems is due to the fact that all their layers exhibit half-metallic FM properties, which suggests a magnetic phase transition emerging from the LMO region~\cite{Cossu2022,Cossu2024}. 
This speculation is confirmed by our calculations of the interatomic exchange interactions, which are found to be stronger (weaker) inside the LMO (SMO) region. The orbital decomposition of the exchange coupling shows that the FM interaction arises from the $E_g$-$E_g$ component and is driven by the double-exchange mechanism. This leading term is weakened by the AFM superexchange involving $t_{2g}$ electrons, which reaches its maximum inside the SMO region. The fact that the total exchange coupling becomes small inside the SMO region introduces a two-phase behavior in the finite-temperature magnetism investigated via ASD simulations. While the thinnest superlattices exhibit a more traditional Curie-Weiss behavior, the superlattice with $n=6$ displays different phase transitions due to distinct regions which order at different temperatures, rather than as a single cohesive unit. 

The calculated Curie temperatures yield values above 450~K for all systems, which seems too large for the thicker systems given their calculated exchange couplings. By revising these values using a correction based on an expected electronic, magnetic, and structural phase transition in the SMO region, we obtain more realistic estimates of about 470~K for $n=2$, 380~K for $n=4$ and 60~K for $n=6$. For comparison, the critical temperature of LSMO is predicted by theory to be around 300~K~\cite{Bottcher_2013}, while experimental values are often slightly higher, likely due to extrinsic effects~\cite{PhysRevLett.104.167203,moussa2025acs}. Superlattices of the same type we investigated here but in the ultrathin limit $n=1$ were recently grown on SrTiO$_3$(111), giving $T_c \approx 320$ K~\cite{wang2022pss}.
While the $n=1$ and $n=2$ structures are not directly equivalent, as the former has a larger cation-intermixing and not enough thickness to accommodate the energetically favorable tilt pattern, the difference between $T_c \approx 320$~K for $n=1$ and our value of 470~K for $n=2$ may be used to define a conservative, worst-case bound on the accuracy of our calculated interatomic exchange interactions. Assuming a roughly proportional scaling between $T_C$ and the overall exchange scale, this would correspond to an overestimation on the order of circa 30\%,which does not affect the overall conclusions of the present study. Therefore, even under this worst-case scenario, we conclude that our theoretical model suggests that (111)-oriented (LMO)$_{2n}$|(SMO)$_n$ superlattices for $n=2,4$ may exhibit robust spin-polarized transport  at a sufficiently high temperature to be useful for spintronic devices.

\vspace{1cm}
\section*{Acknowledgments}
We are thankful to C.\ Autieri, S.\ Sarkar, and V.\ K.\ Lazarov for valuable discussions.
We acknowledge Polish high-performance computing infrastructure PLGrid for awarding this project access to the LUMI supercomputer, owned by the EuroHPC Joint Undertaking, hosted by CSC (Finland) and the LUMI consortium through PLL/2023/04/016450. Additional computational work was performed on resources provided by the National Academic Infrastructure for Supercomputing in Sweden (NAISS), partially funded by the Swedish Research Council through Grant Agreement No. 2022-06725. 
This research is part of the Project No. 2022/45/P/ST3/04247 co-funded by the National Science Center of Poland and the European Union's Horizon 2020 research and innovation program under the Marie Skodowksa-Curie Grant Agreement No. 945339. 
F.\ C. and H.-S.\ K. acknowledge financial support from the National Research Foundation (NRF) funded by the Ministry of Science of Korea (Grants No.\ 2022R1I1A1A01071974). 
The present project was also supported by the STINT Mobility Grant for Internationalization (Grant No. MG2022-9386).

\end{document}


\title{Supplemental material for the manuscript entitled\\ ``Exchange interactions and finite-temperature magnetism in (111)-oriented (LaMnO$_3$)$_{2n}$|(SrMnO$_3$)$_n$ superlattices''
}
\author{Shivalika Sharma}
\email{shivalika@kentech.ac.kr}
\affiliation{Institute of Physics, Nicolaus Copernicus University, 87-100 Toru\'n, Poland}
\author{Julio do Nascimento}
\affiliation{School of Physics, Engineering and Technology, University of York, York YO10 5DD, UK}
\author{Imran Ahamed}
\affiliation{Institute of Physics, Nicolaus Copernicus University, 87-100 Toru\'n, Poland}
\author{Fabrizio Cossu}
\affiliation{School of Physics, Engineering and Technology, University of York, York YO10 5DD, UK}
\affiliation{Department of Physics and Institute of Quantum Convergence and Technology, Kangwon National University, Chuncheon, 24341, Republic of Korea}
\author{Heung-Sik Kim}
\affiliation{Department of Energy Technology, Korea Institute of Energy Technology (KENTECH), Naju, South Korea}
\affiliation{Department of Physics and Institute of Quantum Convergence and Technology, Kangwon National University, Chuncheon, 24341, Republic of Korea}
\author{Igor {Di Marco}}
\email{igor.dimarco@physics.uu.se}
\email{igor.dimarco@umk.pl}
\affiliation{Institute of Physics, Nicolaus Copernicus University, 87-100 Toru\'n, Poland}\affiliation{Department of Physics and Astronomy, Uppsala University, Uppsala 751 20, Sweden}
\date{\today}

\begin{abstract}
In this Supplemental Material, we include layer-resolved spectral properties, selected Mn--O--Mn bond angles as well as a comparison of different projection methods used to evaluate the interatomic exchange interactions.
\end{abstract}
\maketitle

\beginsupplement

\newpage
\section{Layer-resolved spectral properties}
\begin{figure}[t!]
    \centering
    \includegraphics[width=1\linewidth]{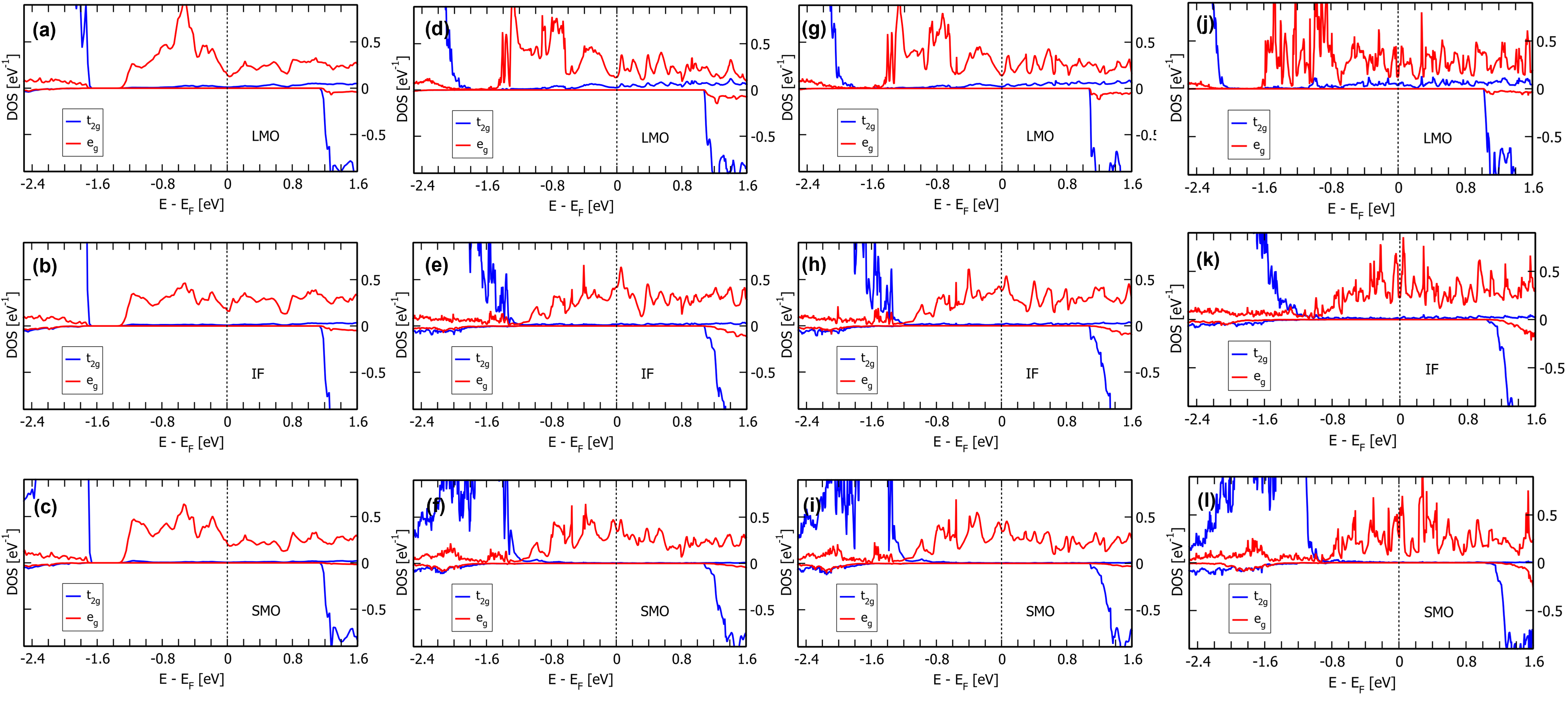}
    \caption{\label{fig:SMpdos}Projected density of states (PDOS) of Mn-$3d$ states in (111)-oriented (LaMnO$_3$)$_{2n}\vert$(SrMnO$_3$)$_{n}$ superlattices in their ferromagnetic (FM) ground state. Data for $n=2$ (a-c), $n=4$  (d-f) and $n = 6$ (j-l) with the $a^-a^-a^-$ tilt system are shown, alongside data for $n=4$ (g-i) with the $a^-a^-c^+$ tilt system. For all systems, we show Mn atoms in the central layer of the LaMnO$_3$ region (LMO), the layer at the interface (IF) and the central layer of the SrMnO$_3$ region (SMO). The Fermi energy is at zero energy.}
\end{figure}
In Figure~\ref{fig:SMpdos}, we show the layer-resolved PDOS for selected Mn atoms in (111)-oriented (LaMnO$_3$)$_{2n}\vert$(SrMnO$_3$)$_{n}$ superlattices for $n=2,4,6$ with $a^-a^-a^-$ tilt system as well as for $n=4$ with the $a^-a^-c^+$ tilt system. All calculations are referred to the FM ground state and are obtained with the Vienna \textit{ab-initio} simulation package (VASP).

\newpage
\section{Comparison of different projectors schemes for the $J_{ij}$}
\begin{figure}[t!]
 \centering
 \includegraphics[width=1\linewidth]{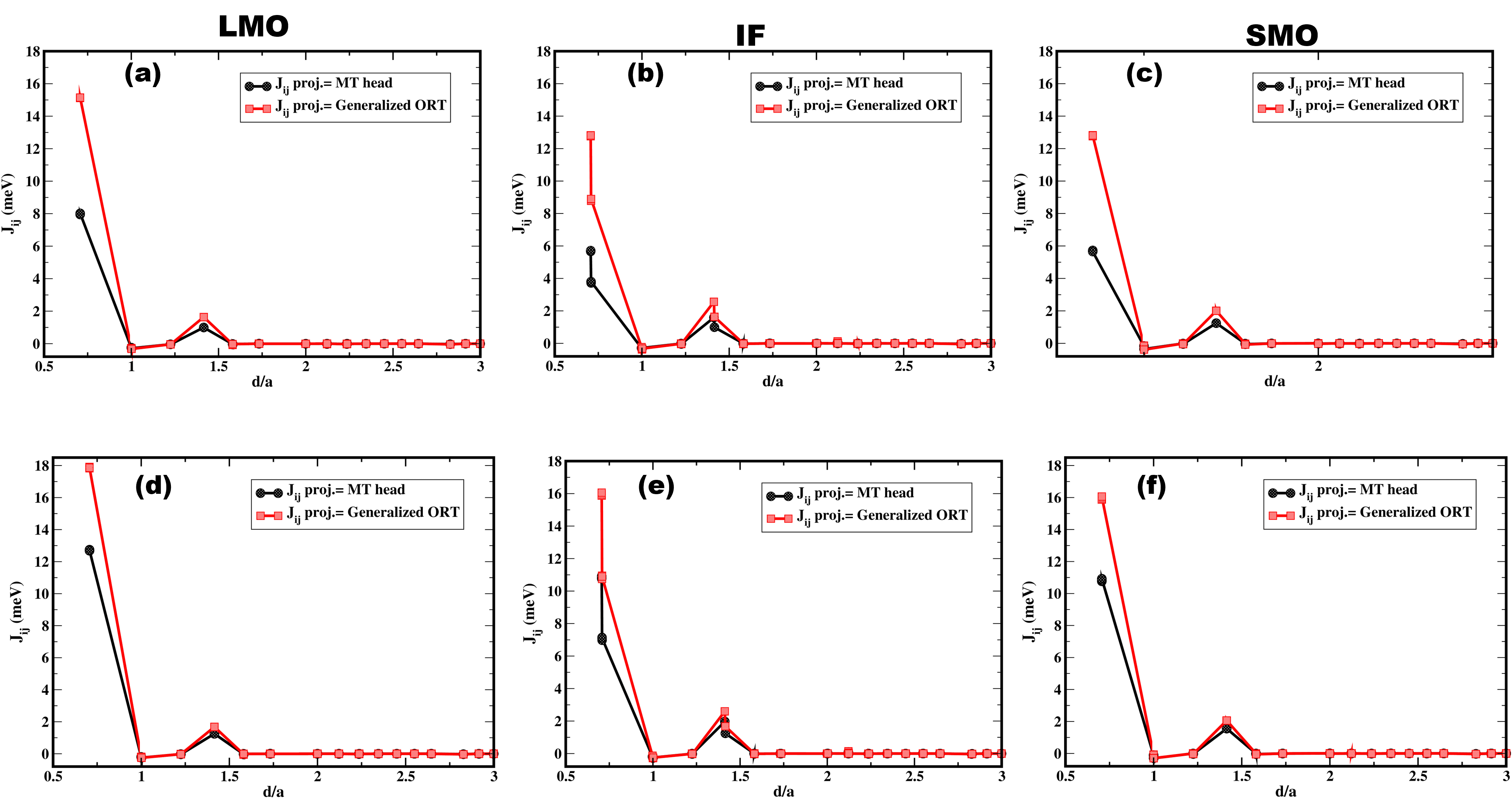} 
 \caption{\label{fig:SMexchange}Comparison of the exchange interaction parameters $J_{ij}$ between Mn magnetic moments at sites $i$ and $j$, as obtained with different projection schemes to evaluate the intersite Green's functions. The black symbols correspond to the muffin-tin (MT) head projectors, while the red symbols correspond to the generalized orthonormalized  (ORT) L\"owdin projectors described in Ref.~\cite{Igor_PhysRevB_2015}. Results are shown for the (111)-oriented (LaMnO$_3$)$_{4}\vert$(SrMnO$_3$)$_{2}$ superlattice in its FM ground state with the $a^-a^-a^-$ tilt system. 
 Plots show data obtained for $U = 3.8$~eV and $J = 1$~eV (a-c) and $U = 5.0$~eV and $J = 1$~eV (d-f). Mn atoms in selected layers are shown, according to the same definitions given in the caption of Figure~\ref{fig:SMpdos}.}
\end{figure}
In RSPt~\cite{RSPt,Igor_PhysRevB_2015}, the interatomic exchange parameters  $J_{ij}$ can be evaluated by means of two different projector schemes for the intersite Green's functions (not to be confused with the local basis defining the Mn-$3d$ states). These schemes usually lead to comparable results, but the usage of only two kinetic energy tails in the RSPt basis set may make the MT projector scheme less accurate in this case, leading a larger discrepancy~\cite{Igor_PhysRevB_2015}. In Figure~\ref{fig:SMexchange}, we compare the $J_{ij}$ between Mn magnetic moments at sites $i$ and $j$ in the investigated superlattices, as obtained from RSPt with different projector schemes. The two trends are qualitatively the same, but ORT gives first nearest-neighbors values that are circa 2 times larger than those given by MT. This difference is much smaller for the fourth nearest neighbors. 

\newpage
\section{Mn--O--Mn bond angles}
\begin{table}[t!]
\caption{Mn--O--Mn bond angles (deg) for LMO, SMO and IF regions in the (111)-oriented (LaMnO$_3$)$_{2n}\vert$(SrMnO$_3$)$_{n}$ superlattices for $n=2,4,6$ in their FM ground state, with the $a^-a^-a^-$ tilt system.}
\label{tab:SMangles}
\centering
\renewcommand{\arraystretch}{1.15}
\begin{tabular}{|c|c|c|c|}
\hline & \multicolumn{3}{c|}{Mn--O--Mn angle (deg)} \\
\hline
  {Layer} & $n=2$ & $n=4$ & $n=6$ \\
\hline
 LMO & \textit{163.9} & \textit{162.1} & \textit{161.2} \\
 IF$_L$  & \textit{163.7} & \textit{164.0} & \textit{164.3} \\
IF$_S$  & \textit{168.6} & \textit{167.8} & \textit{167.5}\\
SMO & \textit{168.6} & \textit{171.2} & \textit{171.8} \\
\hline
\end{tabular}
\end{table}
Table~\ref{tab:SMangles} shows the Mn--O--Mn bond angles obtained for Mn atoms inside LMO, SMO and IF regions, which are defined in the caption of  Figure~\ref{fig:SMpdos}, as calculated for the (111)-oriented (LaMnO$_3$)$_{2n}\vert$(SrMnO$_3$)$_{n}$ superlattices with $n=2,4,6$ in their FM ground state, with the $a^-a^-a^-$ tilt system. For the Mn atom sitting at the IF there are two values, corresponding to the coupling with Mn atoms inside the LMO and SMO regions, and thus labeled as IF$_\text{L}$ and IF$_\text{S}$, respectively. Notice that the angles of the bonds follow the general trend $\mathrm{SMO} > \mathrm{IF}_\mathrm{S} > \mathrm{IF}_\mathrm{L} > \mathrm{LMO}$,
but $\mathrm{IF}_\mathrm{L}$ and LMO are inverted for $n=2$ due to the very small thickness. 

%